\documentclass[%
 reprint,
 amssymb,
 prl,
 superscriptaddress,
 showpacs
]{revtex4-2}
\usepackage{graphicx}
\usepackage{dcolumn}
\usepackage{bm}
\usepackage{amsmath,latexsym}
\usepackage{csquotes}
\usepackage{siunitx}
\usepackage[english]{babel}

\usepackage{float}

\usepackage{braket}
\usepackage[normalem]{ulem}
\usepackage[caption=false, justification=centerlast]{subfig}
\usepackage{todonotes}

\usepackage[%
colorlinks=true,
urlcolor=blue,
linkcolor=blue,
citecolor=blue
]{hyperref}

\newcommand{\UOL}{Department of Physics, The University of Liverpool, Liverpool, L69 3BX, United Kingdom}
\newcommand{\CI}{Cockcroft Institute, Warrington WA4 4AD, United Kingdom}

\newcommand{\UOM}{Department of Physics and Astronomy, The University of Manchester, Manchester M13 9PL, United Kingdom}

\newcommand{\HIJ}{Helmholtz Institute Jena, Fr\"{o}belstieg 3, 07743 Jena, Germany}

\newcommand{\FSU}{Faculty of Physics and Astronomy, Friedrich-Schiller-Universit\"{a}t Jena, 07743 Jena, Germany}

\newcommand{\ICMUV}{ICMUV, Universidad de Valencia, 46071 Valencia, Spain}

\newcommand{\FUHC}{Federal University of Health Sciences of Porto Alegre, Porto Alegre, RS, 90050-170, Brazil}

\newcommand{\GILPA}{Guangdong Institute of Laser Plasma Accelerator Technology, Guangzhou, 510415, China}

\begin{document}
	      

\title{Resonant Excitation of Surface Plasmon for Wakefield Acceleration by Beating GW Lasers on Smooth Cylindrical Surface}

\author{Bifeng Lei}%
\email{bifeng.lei@liverpool.ac.uk}
\affiliation{\UOL}
\affiliation{\CI}

%

\author{Hao Zhang}
\affiliation{\UOL}
\affiliation{\CI}
\author{Alexandre Bonatto}
\affiliation{\FUHC}

\author{Bin Liu}
\affiliation{\GILPA}

\author{Javier Resta‐L\'{o}pez}
\affiliation{\ICMUV}

\author{Matt Zepf}%
\affiliation{\HIJ}
\affiliation{\FSU}

\author{Guoxing Xia}
\affiliation{\UOM}
\affiliation{\CI}
\author{Carsten Welsch}
\affiliation{\UOL}
\affiliation{\CI}

\date{\today}

\begin{abstract}
We present a theoretical and numerical study of resonant surface-plasmon (SP) excitation driven by the beating of two co-propagating laser pulses on a smooth cylindrical plasma–vacuum interface. 
Analytical expressions for the SP dispersion relation, field amplitude, geometric coupling factor, and resonance conditions are derived and validated by fully three-dimensional particle-in-cell simulations. 
We reveal that curvature-induced geometric effects can substantially modify the SP dispersion and enable resonant matching by laser beat waves. This is inaccessible in planar geometries or with a single laser. 
Under matched resonance conditions, a high-amplitude SP-based wakefield can be generated by a few gigawatt (GW) lasers, placing this mechanism within reach of state-of-the-art fibre lasers.
It therefore opens a route toward portable laser-driven plasma wakefield accelerators.
\end{abstract}

\pacs{Valid PACS appear here}
\maketitle



Surface plasmons (SPs) are collective electron oscillations localised at the interface between a conductor and an insulator~\cite{Ritchie:1957aa}. In conventional configurations, SP excitation requires momentum-matching techniques to bridge the mismatch between photon and plasmon wavevectors~\cite{Wood:1902aa, Fano:1941aa, Otto:1968aa}, which becomes increasingly restrictive for high-intensity drivers and limits operational flexibility.

Recent progress in high-power laser-driven grate-free surface plasmons (SPs) on a smooth surface has opened a new route toward ultra-high gradient, ultra-compact plasma-based particle acceleration and radiation generation~\cite{Shen:2021aa, Sarma:2022aa, McCay:2025aa, BFleiCSR_2025, Lei:2025aa}.
In this approach, SPs are excited when the translational symmetry is broken during a laser pulse scattering with a smooth surface of a finite edge~\cite{Stegeman:1983aa}. The efficiency depends primarily on the mode overlap between the incident laser field and the SP eigenfield. 
As surface geometry strongly shapes the SP eigenmodes, it enables the precise mode selection and controlled manipulation. 
 In particular, cylindrical surfaces, such as the inner wall of a microtube, have been shown to offer strong potential for SP-based wakefield acceleration and electron-beam modulation~\cite{BFleiCSR_2025}. 
In the linear regime driven by a weak laser pulse, electrons are confined on the surface, and the SP field can leak into the vacuum channel to form the leaky wakefield, capable of accelerating both positively and negatively charged particles~\cite{Lei:2025aa}. 
In the highly nonlinear regime, SPs can in principle sustain bubble wakefields approaching hundreds of $\si{TV/m}$ inside nanostructured vertically aligned carbon nanotubes (VCNTs)~\cite{Lei:2025ab}. VCNT targets not only offer favourable thermal and electrical properties for efficient SP excitation, but also provide substantial flexibility in achievable plasma density, spanning $10^{20}$--$10^{24}~\si{cm^{-3}}$~\cite{Yu:2009aa}.
However, these concepts typically rely on high-power lasers, ranging from hundreds of terawatts (TW) to petawatts (PW), which remain costly and bulky for many laboratories, and are poorly suited to applications where flexibility and portability are important. Moreover, target survivability under strong electromagnetic (EM) fields of high-power lasers often constrains operation in an effectively single-shot mode, severely limiting the repetition rate.
Low-power lasers can be more compact, economical and comparatively target-friendly, such as fibre lasers, which can deliver up to $5$ gigawatts ($\si{GW}$) power on target~\cite{Lekosiotis:2023aa}.
This power is generally insufficient to directly drive a wakefield of amplitude high enough for electron self-injection and acceleration inside the gaseous bulk plasma~\cite{Esarey:2009aa}. Beating two laser pulses can resonantly enhance the wakefield, which has been explored previously in underdense uniform plasmas~\cite{Kitagawa:1992aa} and on a magnetised planar surface for THz generation~\cite{Chamoli:2024aa}.
However, a viable approach is lacking to use ultralow-power lasers to drive strong SP for relativistic particle acceleration in ultra-compact form.

In this Letter, we demonstrate that SP–based wakefields with amplitudes up to a few $\si{TV/m}$ can be resonantly generated by the beating of two GW-level laser pulses on a smooth cylindrical microtube surface, enabling electron self-trapping and acceleration.
We develop a self-consistent theoretical framework that yields analytical expressions for the SP dispersion relation, field amplitude, geometric coupling factor, and resonance conditions, which are quantitatively validated by fully 3D particle-in-cell (PIC) simulations.
We show that curvature-induced geometric effects substantially modify the SP dispersion, thereby enabling resonant phase matching with laser beat waves, which is inaccessible in planar geometries. Under matched resonance conditions, low-power lasers can efficiently drive high-amplitude SP-based wakefields.
Our results further indicate that laser powers of only a few GW are sufficient to generate SP fields at the tens of $\si{GV/m}$ level, identifying fibre lasers as practical drivers. This work thus opens a pathway toward microscale electron accelerators for portable applications, such as endoscopic sources~\cite{Roa:2022aa}.


\begin{figure}
	\includegraphics[width=0.49\textwidth]{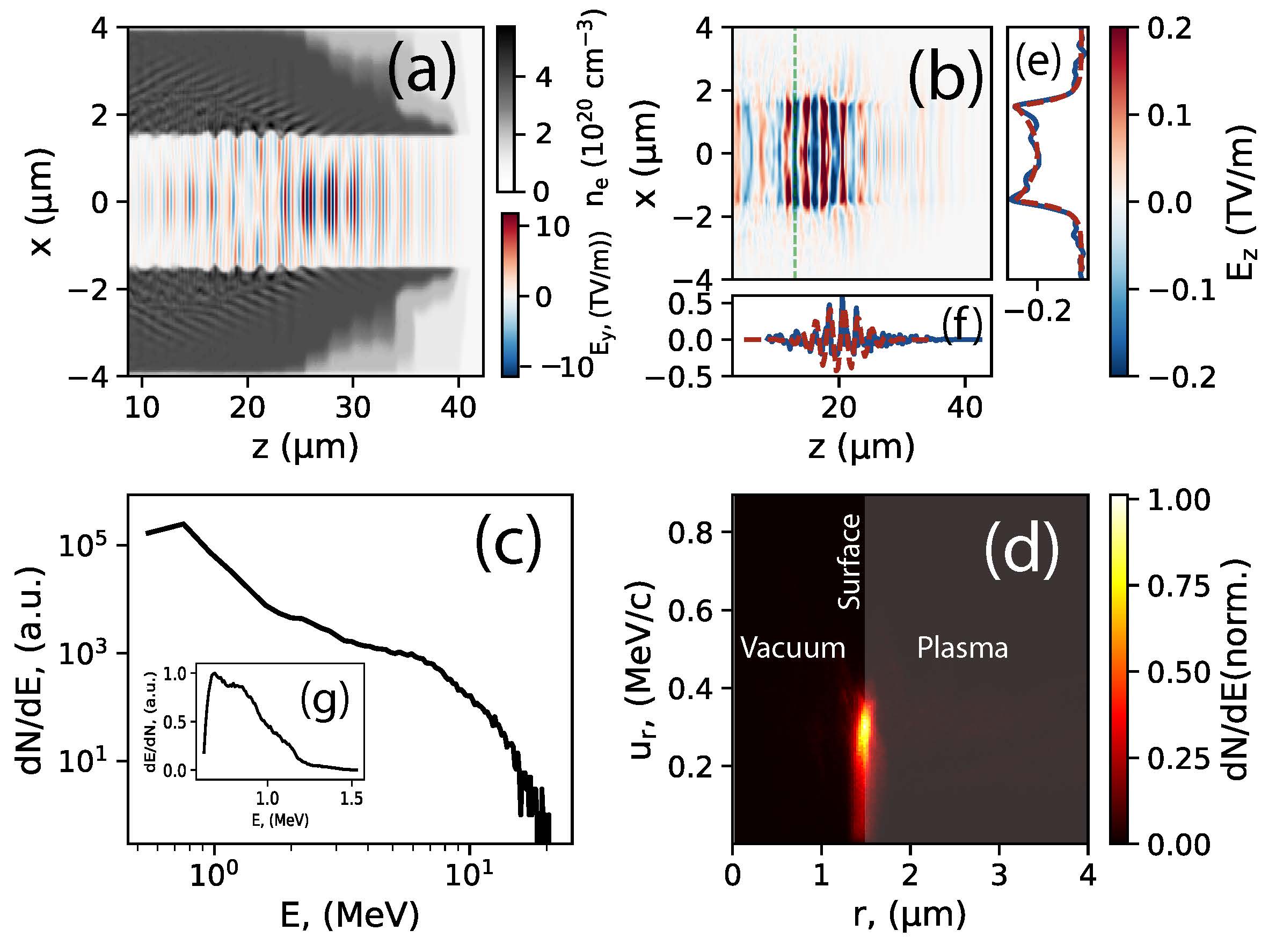}
	\caption{PIC results after two laser pulses co-propagating $40~\si{\mu m}$ inside a solid microtube: (a) electron density (grey) and beating laser field $E_y$ (blue-red) in $xz$ plane.
	(b) Acceleration field $E_z$ in $xz$ plane.
	(c) Energy spectrum of the electron beam accelerated. 
	(d) radial phase space ($r, u_r$) of the accelerated electron beam. The vacuum and plasma regimes are on the left and right, respectively.
	(e) and (f) show the transverse (indicated by green dashed line in (b)) and longitudinal line plots of on-axis ($x=0~\si{\mu m}$) $E_z$. Black solid lines are from PIC, and orange dashed lines are from analytical calculation with $\varrho=1.05$.
	(g) Energy spectrum of the electron beam accelerated inside the vacuum channel.}
	\label{fig:pic_SP_wakefield}
\end{figure}

Beat-wave-driven SP excitation on a smooth cylindrical surface can be demonstrated by a fully 3D PIC simulation done by WarpX~\cite{Vay:2018aa} as shown in Fig.~\ref{fig:pic_SP_wakefield}. 
Two laser pulses of wavelengths $\lambda_1 = 0.8~\si{\mu m}$ and $\lambda_2 = 0.6~\si{\mu m}$ copropagate inside a microtube of a vacuum channel radius $a=1.5~\si{\mu m}$. The normalised laser strength $a_{0,1}=a_{0,2}=0.8$ and the root-mean-square (RMS) pulse durations are the same, $\tau_L = 50~\si{fs}$. These parameters give the laser peak power of $200~\si{GW}$ and $320~\si{GW}$, respectively.  
The carbon atom density is $1 \times 10^{20}~\si{cm^{-3}}$, which can be structured by VCNT forest with a tube density of $1\times 10^{11}~\si{tube/cm^{2}}$ giving the free electron density about  $4 \times 10^{20}~\si{cm^{-3}}$, as such laser pulses can fully ionise the $\sigma$-bond of the carbon atom. The ionisation energies are modified with the CNT properties in simulations. 
The RMS laser spot size is $w_0=3.0~\si{\mu m}$, which gives $60\%$ of on-wall intensity. 
The simulation window is $8~\si{\mu m} \times 8~\si{\mu m} \times 40~\si{\mu m}$ with $192 \times 192 \times 768$ cells in $x$, $y$ and $z$ directions, respectively. There are $2 \times 2 \times 2$ microparticles in each cell. 
Two vacuum sections of $20~\si{\mu m}$ are placed longitudinally at the head and tail of the target for laser initialisation and electron beam diagnostics, respectively.
See more details in the Supplementary Materials.

The laser pulses can stably beat inside the channel and drive the SP as shown in Fig.~\ref{fig:pic_SP_wakefield} (a) and (b), where the structure of the wakefield is capable of accelerating both positively and negatively charged particles. The longitudinal wakefield $E_z$ of amplitude $0.3~\si{TV/m}$ can be resonantly generated inside the vacuum channel, which accelerates totally $0.2~\si{nC}$ electrons to the energy of up to $10~\si{MeV}$ over a $40~\si{\mu m}$ interaction length, as shown in Fig.~\ref{fig:pic_SP_wakefield}(c). This indicates an averaged peak acceleration gradient of $G_{\rm peak}\approx 0.25 ~\si{TeV/m}$. 
While these electrons are confined and accelerated on the surface, the majority of them can propagate longitudinally from the channel due to the strong $E_z$ field. 
There is also a small fraction of electrons, e.g. about $20~\si{fC}$ from this simulation, which can escape transversely from the surface to be trapped during the laser scattering with the vertical edge of the microtube~\cite{Lei:2025aa} and accelerated by the wakefield inside the vacuum channel to produce a narrow energy spectrum as shown in Fig.~\ref{fig:pic_SP_wakefield}(g). The energy spectrum is peaked at $0.8~\si{MeV}$, which indicates the mean acceleration gradient of $G_{\rm mean} \simeq 20~\si{GeV/m}$. 
This is a general picture of the beating wave drive SP and its wakefield acceleration.

SP-based wakefield excitation can be understood as a stimulated Raman scattering (SRS) process~\cite{Antonsen:1992aa}. At the same time, the beat wave-driven SP is not identical to conventional bulk SRS. This interaction can be self-consistently described using Maxwell’s equations coupled to the cold-fluid model for the plasma electrons, formulated in cylindrical coordinates $(r, \phi, z)$.
Inside the vacuum channel, $0\leq r < a$, the permittivity is $\epsilon_v=1$. In the plasma wall, $r \geq a$, $\epsilon_p (\omega)=1-\omega_p^2/ [\gamma_0\omega (\omega+i\nu)]$ where $\omega_p^2=4 \pi n_0 e^2/m_e$. $n_0$ is the equilibrium density of cold electron plasma. $\gamma_0$ is the effective relativistic factor of the electron fluid from the quiver or drift. $\nu$ is an effective collision or damping rate.
Here, we seek axisymmetric modes with a predominantly longitudinal accelerating field. The possible mode is the fundamental TM mode, TM$_0$, which can be driven by ponderomotive~\cite{Lei:2025aa} or space charge force~\cite{Lei:2025ab}. The electromagnetic (EM) fields can be written as $\bm{E}=(E_r,0,E_z)$ and $\bm{B}=(0,B_{\phi}, 0)$. 

For the TM$_0$ mode with $e^{i(k z-\omega t)}$ dependence, the eigen wave equation for $E_z$ is given as~\cite{BFleiCSR_2025}
\begin{equation}
	\left ( \frac{\partial^2}{\partial r^2} + \frac{1}{r} \frac{\partial}{\partial r} \right) E_z + \left( \epsilon_j \frac{\omega^2}{c^2} - k^2 \right ) E_z = 0 \mathrm{,}
	\label{eq:wave_eq_Ez}
\end{equation} 
where $j$ denotes $v$ or $p$, representing the equation in vacuum or plasma. $k$ and $\omega$ represent the propagation constant and oscillation frequency of the mode, respectively. Therefore, we can define the radial evanescent constants as $\kappa_v=\sqrt{k^2 - \omega^2/c^2}$ and $\kappa_p=\sqrt{k^2 - \epsilon_p (\omega)\omega^2/c^2}$.
The solution is given inside the vacuum as $E_z^{(v)}(r,z,t)=A I_0(\kappa_v r) e^{i(kz - \omega t)}+c.c.$ and inside the plasma as $E_z^{(p)}(r,z,t)=C K_0(\kappa_p r) e^{i(kz - \omega t)}+c.c.$, where $I_0$ and $K_0$ are modified Bessel functions of first and second kinds, respectively. $A$ is a constant and $C=AI_0(\kappa_v a)/K_0(\kappa_pa)$. The radial electric field $E_r^{(j)}$ and azimuthal magnetic field $B_{\phi}^{(j)}$ are given as $E_r^{(j)} = (ik/\kappa_j^2) d E_z^{(j)} / d r$ and $B_{\phi}^{(j)}=(i \omega \epsilon_j k /\kappa_j^2) d E_z^{(j)}/d r$.

\begin{figure}
 	\centering
 	\includegraphics[width=0.45\textwidth]{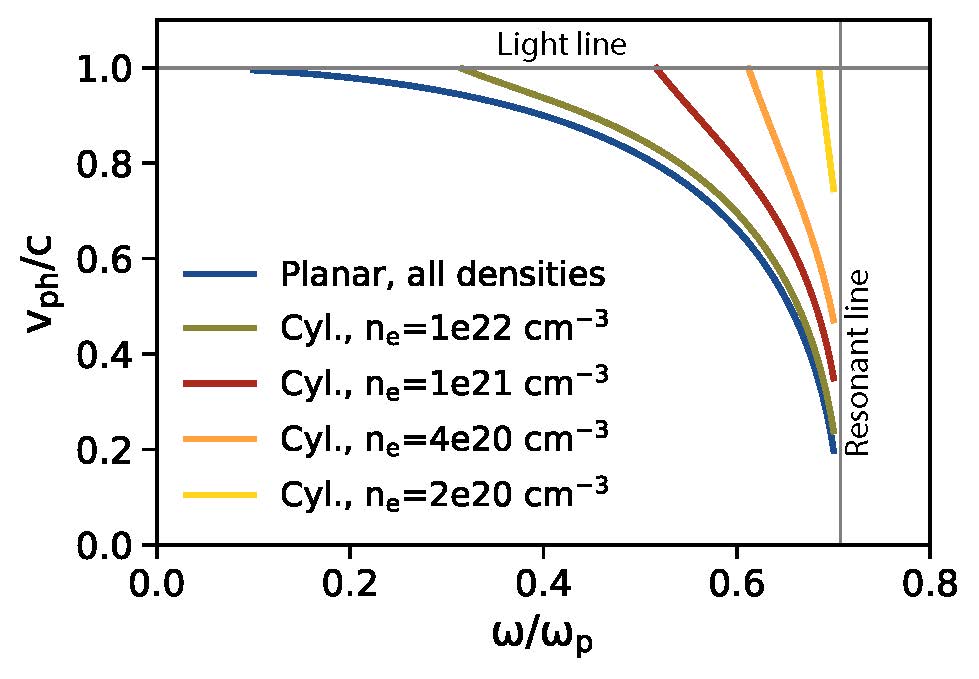}
 	\caption{Phase velocity $v_{\rm ph}$ of SP on planar and cylindrical surfaces of different plasma density, calculated with $a=1.0~\si{\mu m}$ by Eq.~\eqref{eq:disper_relation}.}
 	\label{fig:phase_velocity}
\end{figure}

The dispersion function is defined by considering the boundary conditions at the interface $r=a$ where tangential $E_z$ and $B_{\phi}$ are continuous as $E_{z}^{(v)}(a)=E_{z}^{(p)}(a)$ and $B_{\phi}^{(v)}(a)=B_{\phi}^{(p)}(a)$, given by
\begin{equation}
	\mathcal{D} (\omega, k) = \frac{\epsilon_v}{\kappa_v} \frac{I_1(\kappa_v a)}{I_0(\kappa_v a)} + \frac{\epsilon_p}{\kappa_p} \frac{K_1(\kappa_p a)}{K_0(\kappa_p a)} \mathrm{,}
	\label{eq:disper_relation}
\end{equation}
where the eigenmode satisfies $\mathcal{D}(\omega, k)=0$. The phase velocity is $v_{\rm ph}=\omega/k$ and group velocity $v_g=-(\partial \mathcal{D}/\partial k) / (\partial \mathcal{D}/ \partial \omega)$. 
In the large-radius limit $k a \gg 1$ or more precisely $\kappa_j^{-1} \ll a$, the dispersion in Eq.~\eqref{eq:disper_relation} recovers to that on a planar surface with a curvature correction in order of $\mathcal{O}(1/ka)$.
The phase velocity is shown in Fig.~\ref{fig:phase_velocity}, where this curvature correction leads to the divergence for different plasma densities. In contrast, all $v_{\rm ph}/c$ collapse onto the same curve on planar surfaces. This is the fundamental reason for the advantage of curved surfaces, as discussed later.
Generally, in the vicinity of the SP resonance, $\omega/\omega_p \to 1/\sqrt{2}$, the wavenumber diverges $k \to \infty$, while both $v_{\rm ph}$ and $v_{\rm g}$ vanish as $v_{\rm ph}\to 0$ and $v_g \to 0$, corresponding to a standing wave strongly localized at the plasma–vacuum interface. 
In contrast, for $\omega/\omega_p \ll 1$, the dispersion approaches the light line, with $v_{\rm ph} \approx v_g \approx c$, which indicates a weakly bound SP. Notably, on a cylindrical interface, the phase velocity remains significantly higher near the resonant line than in the planar case due to curvature effects. This enhancement is particularly pronounced for lower plasma densities and therefore allows the resonant SP excitation at a near-critical-density plasma for relativistic particle acceleration.

A physical SP field depends on the specific drive for the eigen wave equation in Eq.~\eqref{eq:wave_eq_Ez}. Let's consider two co-propagating linearly polarised laser pulses of sufficiently long duration as $\bm{a}_{L}(r,z,t)=\Re \left \{ \bm{e}_{\perp} a_{0,1} f_1(r) e^{i(k_1z-\omega_1 t)} + \bm{e}_{\perp} a_{0,2}f_2(r,t) e^{i(k_2 z-\omega_2 t)} \right \}$, where $f_l(r)$ is the laser transverse profile and $a_{0,l}$ is the amplitude of normalised vector potential with $l$ denoting either $1$ or $2$ for the first and second laser pulses, respectively. Here, we consider two laser pulses of the same profiles as $f_l(r)=f(r)$, which does not introduce any new physics.
The frequencies and wavenumbers satisfy $\omega_l\simeq c k_l$ in vacuum.
They create a slowly varying beating intensity of the time-averaged squared amplitude $\langle a_L^2(r,z,t)\rangle = a_{0,1}^2 f^2(r) + a_{0,2}^2 f^2(r) + 2|a_{0,1}| |a_{0,2}| f^2(r)\cos(\Delta k z - \Delta \omega t)$ with $\Delta k=k_1-k_2$ and $\Delta \omega = \omega_1 - \omega_2$. 
The cycle-averaged ponderomotive potential for electrons near the surface is given by $U_p(z,t)\simeq m_e c^2 \langle a_L^2(r,z,t)\rangle/4$. 
The beat ponderomotive force on electrons is $\bm{F}_p^{\rm{beat}}=- \nabla U_p = F_{p,r} \bm{e}_r + F_{p,z}\bm{e}_z$, where the longitudinal component is given by $F_{p,z}(z,t) = [m_e c^2 a_{0,1} a_{0,2} \Delta k  \ \sin(\Delta k z - \Delta \omega t)]/2$. 
For a solid surface and an optical laser beam where $\omega_{p} w_0/c \gg 1$, the longitudinal beat force has much better overlap with the SP eigenmode than the transverse one. Therefore, $F_{p,z}$ dominantly modulates the surface electron density and current at the beat frequency and wavenumber $(\Delta \omega, \Delta k)$. 
The radial force only creates the necessary charge separation across the vacuum-plasma interface to excite the surface mode.
The feature is highly beneficial for maintaining the stability of the interface.
See more discussions in the Supplementary Materials.

The physical field can be written as $\bm{E}^{\text{SP}}(r,z,t) = \Re \left \{ \bm{\mathcal{E}}^{(j)}(r) A(z,t) e^{i (k_{sp}z - \omega_{sp}t)} \right \}$, where $A(z,t)$ is complex amplitude. $\omega_{sp}$ and $k_{sp}$ are the eigen frequency and wavenumber and satisfy the dispersion relation in Eq.~\eqref{eq:disper_relation}.  $\bm{\mathcal{E}}$ is the normalized eigenmode profile along $\hat{z}$-direction as $\bm{\mathcal{E}} = \bm{E}/ \mathcal{N}_{\rm{SP}}$ where $\mathcal{N}_{\rm{SP}}$ is the mode energy as $\mathcal{N}_{\rm{SP}} = \frac{1}{16 \pi} \int \left( |\bm{B}|^2 + |\bm{E}|^2 \right ) d^2 r_{\perp}$ with $\int d^2 r_{\perp}$ denoting the integration over the cross-section transverse to $z$-direction.
Projecting the Maxwell equations onto the eigenmode yields a driven, damped oscillator for $A(z,t)$ as 
\begin{equation}
\begin{split}
		\bigg (\frac{\partial^2}{\partial t^2} & + 2 \nu \frac{\partial }{\partial t} + \omega_{sp}^2 - v_g^2 \frac{\partial^2}{\partial z^2} \bigg ) A(z,t) \\
		& = S_{\rm{SP}} a_{0,1} a_{0,2} \Delta k e^{i(\Delta k z-\Delta \omega t)} \mathrm{,}
\end{split}
\label{eq:A_wave_equ}
\end{equation}
where $S_{\rm{SP}}$ is a coupling coefficient that depends on the overlap between the laser ponderomotive source and the SP eigenmode. It is calculated as
\begin{equation}
\begin{split}
		S_{\rm{SP}} & = \int d^2 r_{\perp} \bm{\mathcal{E}}^{*}(r) \cdot \bm{F}_p^{\rm{beat}}(r) \\
		&\simeq - \frac{i \pi e n_0 c^2}{2 \gamma_0 \omega_{sp}^2} \int_{0}^{\infty} r dr  \mathcal{E}^{*}_z(r) \ f^2(r) \mathrm{,}
		\label{eq:S_SP}
\end{split}
\end{equation} 
where the radial contribution is neglected. See more details in the Supplementary Materials.

For a monochromatic near-resonant drive with slowly varying $A(z,t)$, the steady-state solution is approximately 
\begin{equation}
	A(z,t) \approx - \frac{S_{\rm{SP}} a_{0,1} a_{0,2} \Delta k}{\omega_{sp}^2 - v_g^2 \Delta k^2 - \Delta \omega^2 - 2 i \nu \Delta \omega} e^{i(\Delta k z - \Delta \omega t)} \mathrm{.}
	\label{eq:wake_amp}
\end{equation}
The peak amplitude is then scaled as 
\begin{equation}
	E_{z,\text{max}}^{\rm wake}  \approx G_{\rm{SP}} \ a_{0,1} a_{0,2} \ E_{\rm{SP},0} \mathrm{,}
\end{equation}
where the natural SP field scale is defined by $E_{\rm{SP},0} = m_e c \omega_{sp}/e$.
The dimensionless geometric factor $G_{\rm{SP}}$ measures how efficiently the beat ponderomotive force deposits energy into the normalised SP mode, and is defined by
\begin{equation}
	G_{\rm{SP}} \equiv \frac{|\mathcal{E}_z(a)|}{ E_{\rm{SP,0}}} \frac{|S_{\rm{SP}}|\Delta k}{\nu_{\rm eff}} \mathrm{,} 
\end{equation}
where the effective damping is defined by $\nu_{\rm eff}=\sqrt{(\omega_{sp}^2 - v_g^2 \Delta k^2 - \Delta \omega^2)^2 + (2 \nu \Delta\omega)^2}$.
It is easy to see that $G_{\rm SP}$ carries the information of geometry, damping and dispersion and depends on the material properties. $G_\mathrm{SP}$  is large if 1) $|\mathcal{E}_z(a)|$ is large (e.g. the mode is strongly surface localised); 2) $|S_{\rm SP}|$ is large (e.g. the laser beam is well matched to the mode radial structure); 3) detuning is small (e.g. the beam matches the SP dispersion); 4) $\nu$ is small (e.g. modest damping).
The analytical solution of the SP field agrees very well with the PIC simulation, as shown in Fig.~\ref{fig:pic_SP_wakefield} (e) and (f).

\begin{figure}
	\centering
	\includegraphics[width=0.45\textwidth]{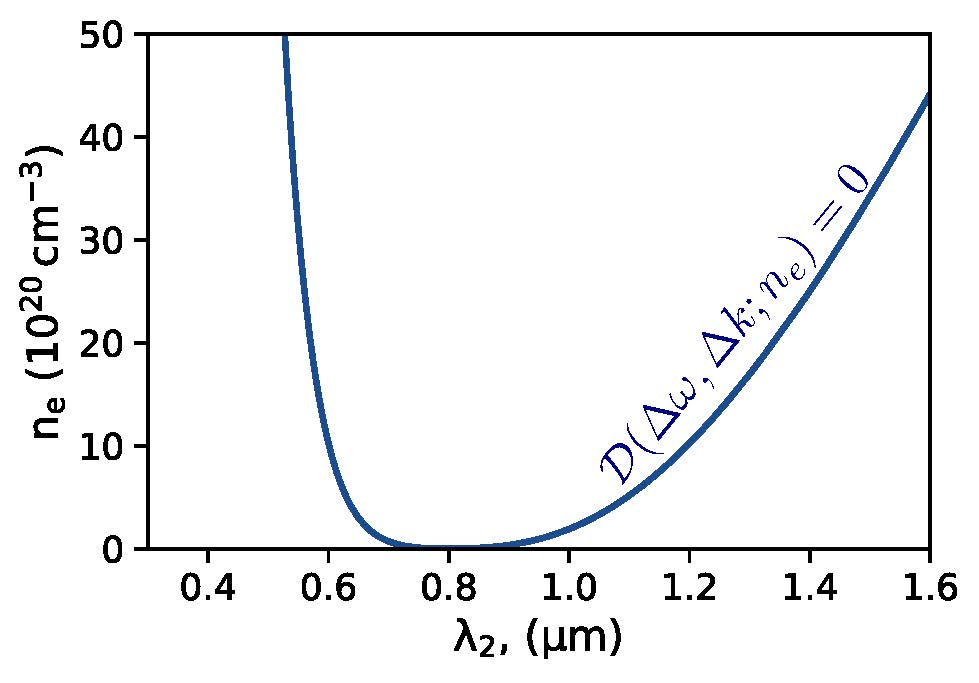}
	\caption{Resonant condition between matched plasma density  $n_e$ and second laser wavelength $\lambda_2$, where $\lambda_1=0.8~\si{\mu m}$, $a=1.5~\si{\mu m}$ and $\gamma_0=1.0$, calculated by Eq.~\eqref{eq:light_resonance}.}
	\label{fig:resonance_condition}
\end{figure}

\begin{figure}
\centering
	\includegraphics[width=0.45\textwidth]{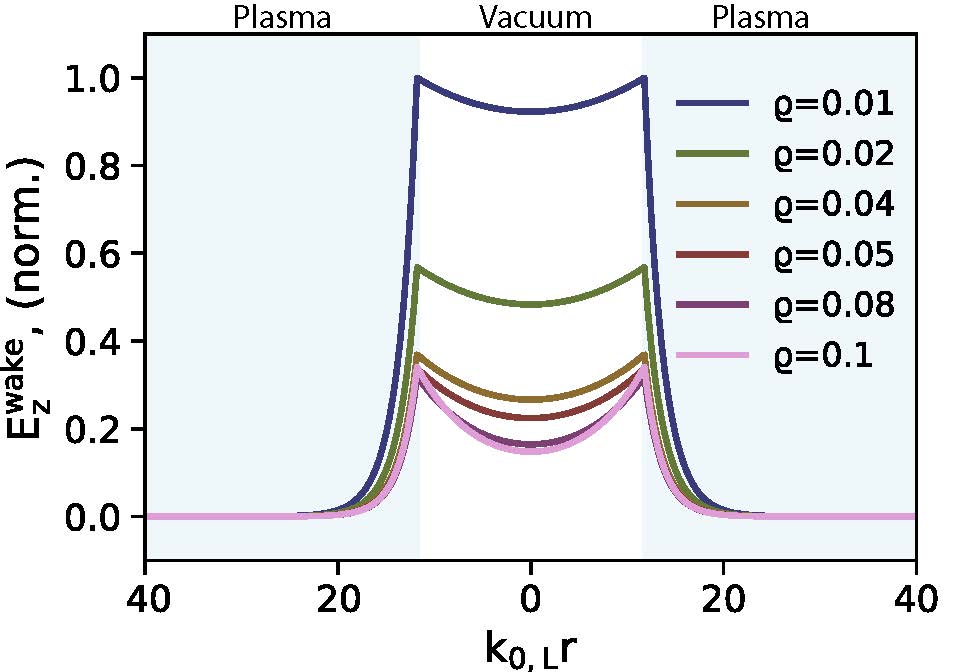}
	\caption{Radial distribution of longitudinal component of electric field $E_z^{\rm wake}$ for different values of detuning rate $\varrho$ defined by $\varrho=(\Delta \omega - \alpha \Delta k)/(\omega_{sp} - \alpha k_{sp})-1$. Here, $\nu=0$. The light blue regions indicate the plasma wall.}
	\label{fig:detuning}
\end{figure}

The resonance condition in Eq.~\eqref{eq:wake_amp} is $\omega_{sp}^2-v_g^2 \Delta k^2 - \Delta \omega^2\approx 0$ for weak damping. 
The exact resonance occurs when $\Delta \omega = \omega_{sp}$ and $\Delta k =k_{sp}$. 
The cylindrical resonant condition is obtained by satisfying the dispersion relation in Eq.~\eqref{eq:disper_relation} as $\mathcal{D}(\Delta \omega, \Delta k; n_e)=0$. This means to choose a matched plasma density such that the cylindrical SP intersects the light line exactly at the beat frequency. 
If the radial decay of the SP field length in the core is much larger than the tube radius, as shown in Fig.~\ref{fig:detuning}, we have $\lim_{\kappa_v a\to 0}  I_1(\kappa_v a)/(\kappa_v I_0(\kappa_v a)) = a/2$. As a result, the cylindrical resonant condition can be reduced to
\begin{equation}
	\frac{1}{2} + \frac{\epsilon_p(\Delta \omega)}{\zeta} \frac{K_1(\zeta)}{K_0(\zeta)} = 0 \mathrm{,} 
	\label{eq:light_resonance}
\end{equation}
where $\zeta = \omega_p a/\gamma_0 c$.
The resonant condition is numerically calculated and shown in Fig.~\ref{fig:resonance_condition}. It is seen that for a finite density, the SP mode lives around $\omega \sim \mathcal{O}(\omega_p)$, and does not extend down to arbitrarily small $\omega$ on the light line where $\Delta \omega \to 0$.
This implies that a single on-axis laser cannot resonantly excite a finite-frequency SP without additional asymmetry.
The most efficient excitation occurs when the cylindrical SP mode admits $v_{\rm ph}$ close to $c$, while the field enhancement benefits from approaching the bound regime near the resonant line. 
The curvature allows the SP to be significantly faster than in the planar case at a similar density.   Eq.~\eqref{eq:light_resonance} implies a $v_{\rm ph}$ close to $c$ while staying in a near-critical regime as shown in Fig.~\ref{fig:phase_velocity}. That’s the key advantage of the cylindrical surface.


Eq.~\eqref{eq:light_resonance} is a joint condition on the lasers and the plasma dispersion and geometry, with a trade-off controlled by detuning and damping.
The field amplitude is significantly limited by detuning on a cold plasma surface, as shown in Fig.~\ref{fig:detuning}.
By linearising the dispersion around $(\omega_{sp}, k_{sp})$ near the resonance, we have $\omega^2 \approx \omega_{sp}^2 + 2 \omega_{sp} \delta \omega$ and $k^2 \approx k_{sp}^2 + 2 k_{sp} \delta k$ with $\delta \omega = \Delta \omega - \omega_{sp}$ and $\delta k = \Delta k - k_{sp}$. The effective damping is reduced to $\nu_{\rm eff}^{\rm res}\simeq 2\omega_{sp} \sqrt{(\delta \omega+\alpha \delta k)^2 + \nu^2}$ with $\alpha=v_g^2k_{sp}/\omega_{sp}^2$. 
This will be a major experimental concern due to the inevitable laser-plasma mismatch in a real experiment. 


\begin{figure}
	\includegraphics[width=0.45\textwidth]{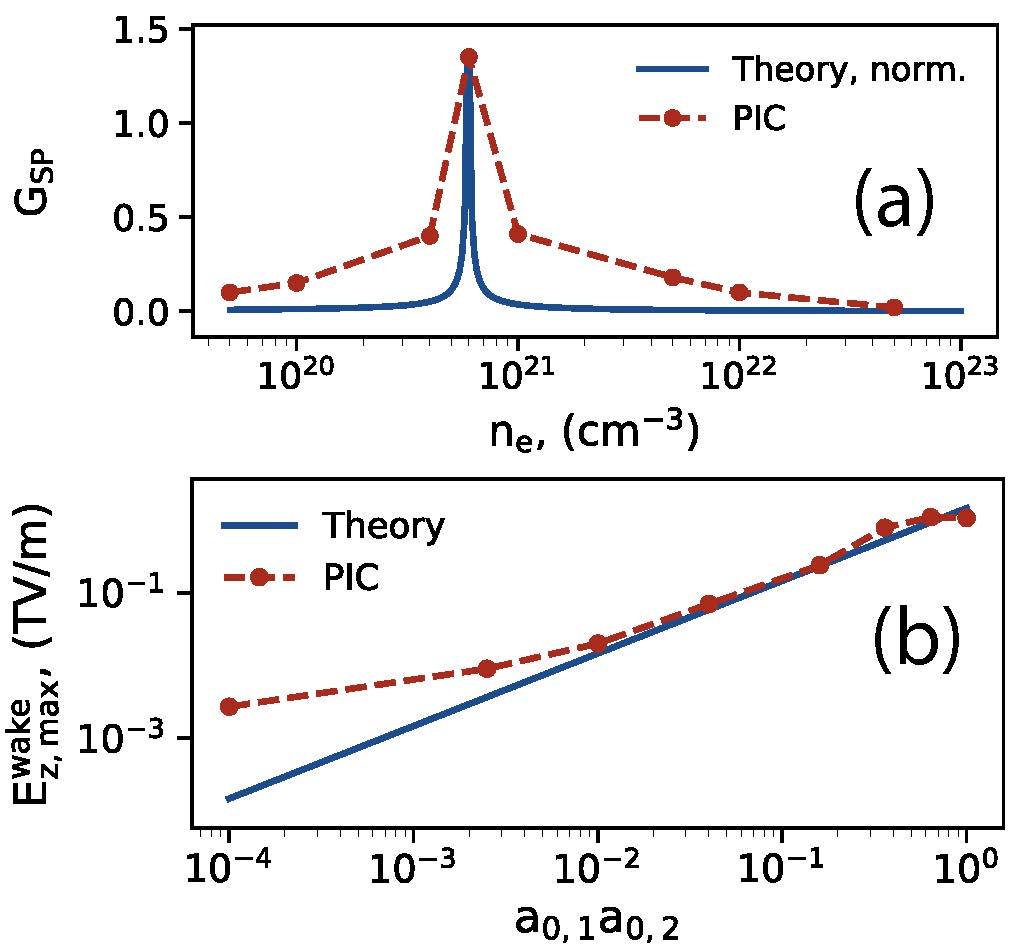}
	\caption{(a) $G_{\rm SP}$ as a function of plasma density $n_e$ from theory (normalised, blue solid) and PICs (red-dot).
	(b) Amplitude of on-axis acceleration field $E_{\rm z, max}^{\rm wake}$ as a function of $a_0^2$ from theory (blue solid) and PICs (red-dot). Here, $a_{0,1}=a_{0,2}=a_0$.
	The other parameters are $\lambda_1=0.8~\si{\mu m}$, $\lambda_2=0.6~\si{\mu m}$, $a=1.0~\si{\mu m}$ and $w_0=2.0~\si{\mu m}$. In (a), and $a_{0}=0.6$, corresponding to laser peak powers $P_{\rm 1, peak}=48~\si{GW}$ and $P_{\rm 2, peak}=80~\si{GW}$, respectively. In (b), $n_e=6\times 10^{20}~\si{cm^{-3}}$, close to the resonant density. The PIC results are obtained after the laser pulses co-propagate $40~\si{\mu m}$ inside a microtube.
	}
	\label{fig:G_SP}
	\end{figure}

In PIC simulations, $G_{\rm SP}$ can be simply calculated by $G_{\rm SP}^{\rm PIC} = E_{z, \mathrm{max}}^{\rm wake}/(a_{0,1} a_{0,2} \ E_{\rm SP, 0})$ and presents the resonant enhancement, while the acceleration field scales with $a_{0,1} a_{0,2}$, agreeing well with the theory, as shown in Fig.~\ref{fig:G_SP}(a).
Resonant SP excitation enables low-power lasers to act as effective drivers, as shown in Fig.~\ref{fig:G_SP}(b), where $a_{0,1}=a_{0,2}=0.01$ corresponds to a $0.15~\si{GW}$ peak power for $\lambda_1=0.8~\si{\mu m}$ and a $0.25~\si{GW}$ for $\lambda_2=0.6~\si{\mu m}$ with a acceleration field of $~\sim 1~\si{GV/m}$.
It can be seen that SP growth is much faster in the low-power regime. This is due to the small detuning rate caused by the weak surface density perturbation.

SP excitation occurs as soon as the outermost electrons of the surface material can be ionised by the laser field.
For example, on a metallic crystalline surface, such as VCNT, laser pulses with a peak power of order $10~\si{MW}$ are sufficient to ionise the first $\pi$-bond electrons, thereby initiating SP resonant growth. However, at such low intensities, the resulting SP field remains too weak to trap electrons. 
Current PIC simulations in Fig.~\ref{fig:G_SP} (b) indicate that efficient electron trapping occurs for $a_0>0.2$, corresponding to a laser power threshold of approximately $4~\si{GW}$. Importantly, this power level lies well within the capabilities of currently available high-power fibre laser systems~\cite{Lekosiotis:2023aa}, demonstrating the experimental feasibility of resonant SP-driven wakefield acceleration.

In summary, the theory of resonant SP excitation driven by beating lasers presented in this letter can be readily generalised to other drivers, including the Lorentz or ponderomotive force of a laser field and the space-charge force of a charged-particle beam. In each case, Eq.~\eqref{eq:A_wave_equ} is modified by the appropriate source term, and the corresponding eigenmodes become dominant depending on the coupling efficiency. Geometric effects can be treated similarly, and the efficiency can be improved by using structured drivers, such as Laguerre-Gaussian (LG) lasers.
While this scheme works with lasers of a wide range of power from $\si{GW}$ to $\si{PW}$, a particularly attractive prospect is to accelerate a well-collimated electron beam up to a few $\si{MeV}$ inside a solid microtube using portable $\si{GW}$ lasers over a 10s of $\si{\mu m}$ scale, while retaining substantial flexibility in energy spectrum, brightness, and beam size.
Technically, this could be realised by attaching an ultrathin film to the output of standard laser optics. Such films can be fabricated with existing techniques using VCNT with well-structured channels.
Therefore, this beat-wave-driven scheme merits consideration as a platform technology with potential impact across materials science, medicine, computing, communications, and energy.

\begin{acknowledgements}
This work has been supported by the Science and Technology Facilities Council (STFC) through the Cockcroft Institute core grant UKRI1887.
Guoxing Xia acknowledges the support from the Cockcroft Institute Core Grant No. ST/V001612/1.
Javier Resta-López acknowledges support by the Generalitat Valenciana under grant agreement CIDEGENT/2019/058.
Bin Liu acknowledges the support of Guangdong Basic and Applied Basic Research Foundation (Grant No. 2025A1515011532) and Guangdong Major Talent Project (Grant No. 2023CX10Y036). 
This work made use of the Barkla High Performance Computing facilities at the University of Liverpool.
\end{acknowledgements}

\

\bibliography{bwSP.bib}

\end{document}